\documentclass[aps,twocolumn,showpacs,preprintnumbers,superscriptaddress,amsmath,amssymb]{revtex4}
\usepackage[pdftex]{graphicx}
\usepackage[english]{babel}
\usepackage[ansinew]{inputenc}
\usepackage{natbib}
\usepackage{color}
\usepackage[normalem]{ulem} 

\newcommand{\ie}{{i.e. }}

\newcommand{\cf}{{cf. }}

\newcommand{\co}[2]{\ifcase #1 \or #2 \fi}
\usepackage{ulem}
\newcommand{\bscco}{Bi$_{2}$Sr$_{2}$CaCu$_{2}$O$_{8}$\,}
\newcommand{\micron}{$\,\mu$m }

\begin{document}
\title{Hot-spot formation in stacks of intrinsic Josephson junctions in Bi$_{2}$Sr$_{2}$CaCu$_{2}$O$_{8}$}
\author{B. Gross}
\affiliation{Physikalisches Institut -- Experimentalphysik II and Center for Collective Quantum Phenomena in LISA$^+$, Universit\"{a}t T\"{u}bingen, Auf der Morgenstelle 14, D-72076 T\"{u}bingen, Germany}
\author{S. Gu\'{e}non}
\affiliation{Physikalisches Institut -- Experimentalphysik II and Center for Collective Quantum Phenomena in LISA$^+$, Universit\"{a}t T\"{u}bingen, Auf der Morgenstelle 14, D-72076 T\"{u}bingen, Germany}
\affiliation{Department of Physics, Center for Advanced Nanoscience, University of California-San Diego, La Jolla, California 92093, USA}
\author{J. Yuan}
\affiliation{National Institute for Materials Science, Tsukuba 3050047, Japan}
\author{M.Y. Li}
\affiliation{National Institute for Materials Science, Tsukuba 3050047, Japan}
\affiliation{Research Institute of Superconductor Electronics, Nanjing University, Nanjing 210093, China}
\author{J.~Li}
\affiliation{National Institute for Materials Science, Tsukuba 3050047, Japan}
\author{A.~Ishii}
\affiliation{National Institute for Materials Science, Tsukuba 3050047, Japan}
\author{R.G.~Mints}
\affiliation{The Raymond and Beverly Sackler School of Physics and Astronomy, Tel Aviv University, Tel Aviv 69978, Israel}
\author{T.~Hatano}
\affiliation{National Institute for Materials Science, Tsukuba 3050047, Japan}
\author{P.H.~Wu}
\affiliation{Research Institute of Superconductor Electronics, Nanjing University, Nanjing 210093, China}
\author{D.~Koelle}
\affiliation{Physikalisches Institut -- Experimentalphysik II and Center for Collective Quantum Phenomena in LISA$^+$, Universit\"{a}t T\"{u}bingen, Auf der Morgenstelle 14, D-72076 T\"{u}bingen, Germany}
\author{H.B.~Wang}
\email{hbwang1000@gmail.com}
\affiliation{National Institute for Materials Science, Tsukuba 3050047, Japan}
\affiliation{Research Institute of Superconductor Electronics, Nanjing University, Nanjing 210093, China}
\author{R.~Kleiner}
\email{kleiner@uni-tuebingen.de}
\affiliation{Physikalisches Institut -- Experimentalphysik II and Center for Collective Quantum Phenomena in LISA$^+$, Universit\"{a}t T\"{u}bingen, Auf der Morgenstelle 14, D-72076 T\"{u}bingen, Germany}%
\date{\today}%
%
\begin{abstract}
We have studied experimentally and numerically temperature profiles and the formation of hot spots in intrinsic Josephson junction stacks in \bscco (BSCCO). The superconducting stacks are biased in a state where all junctions are resistive. The formation of hot spots in this system is shown to arise mainly from the strongly negative temperature coefficient of the c-axis resistivity of BSCCO at low temperatures. This leads to situations where the maximum temperature in the hot spot can be below or above the superconducting transition temperature $T_c$. The numerical simulations are in good agreement with the experimental observations. 

\end{abstract}
\pacs{74.50.+r, 74.72.-h, 85.25.Cp}
%
%
\maketitle
%
%
\section{Introduction}
\label{sec:intro}
Joule heating is an omnipresent issue in current-carrying structures and has been studied for a long time.
General aspects, like the propagation of switching waves or the formation of static electrothermal domains in bistable conductors are well-known phenomena \cite{Gurevich87,Volkov69}. In Josephson junctions heating often is small enough to be neglected. An exception are stacks of intrinsic Josephson junctions (IJJs) in the high temperature superconductor \bscco (BSCCO). Here, the BSCCO crystal structure intrinsically forms stacks of Josephson junctions, each having a thickness of 1.5\,nm. A single IJJ may carry a voltage $V$ of some mV and a current $I$ of several mA. Although the dissipative power generated by a single IJJ is only some $\mu$W, the power inside a stack of, say, 1000 IJJs amounts to several mW, with power densities well in excess of $10^4$ W/cm$^3$. For small sized ($\sim$ a few $\mu$m in diameter, consisting of some 10 IJJs) stacks the corresponding overheating has been discussed intensively in literature \cite{Thomas00,Fenton03, Anagawa03, Zavaritsky04, Yurgens04, Krasnov05, Wang05, Verreet07}.
\begin{figure}
\includegraphics[width=0.8\columnwidth]{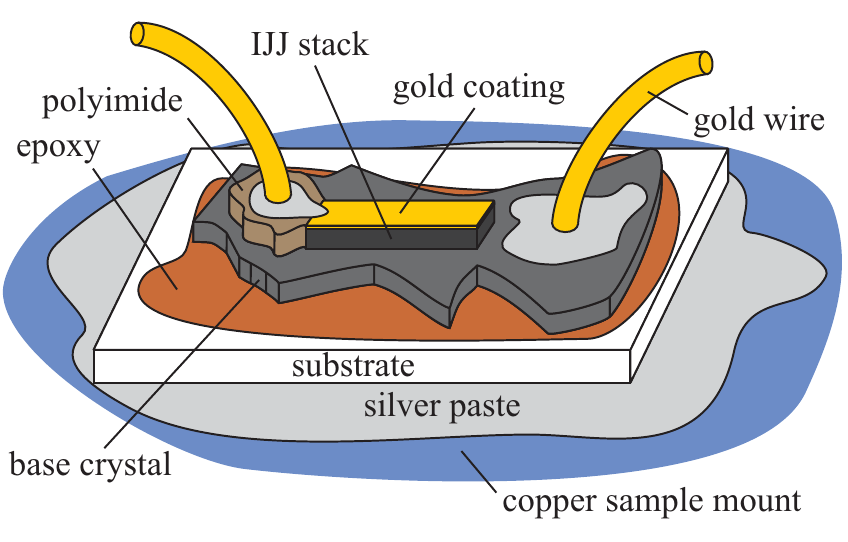}
\caption{(Color online) Typical design of BSCCO IJJ mesas.}
\label{fig:typicalsample}
\end{figure}

Recently, coherent off-chip THz radiation with an extrapolated output power of some $\mu$W was observed from stacks of more than 600 IJJs, with lateral dimensions in the 100$\,\mu$m range \cite{Ozyuzer07}. The IJJ stacks have been patterned in the form of mesa structures, as shown schematically in Fig. \ref{fig:typicalsample}. THz radiation emitted from such IJJ stacks became a hot topic in recent years, both in terms of experiment \cite{Ozyuzer07, Kadowaki08, Ozyuzer09a, Wang09a, Minami09, Kurter09, Gray09, Kadowaki10, Guenon10, Wang10a, Tsujimoto10, Koseoglu11, Benseman11, Yamaki11, Kashiwagi12, Tsujimoto12, Li12}
and theory \cite{Bulaevskii07, Koshelev08, Koshelev08b, Lin08, Krasnov09, Hu09, Nonomura09, Tachiki09, Koyama09, Pedersen09, Klemm10, Krasnov10, Katterwe10, Savelev10, Hu10, TachikiT10, Lin10, Zhou10, Koshelev10, Yurgens10a, Tachiki11, Yurgens11, Yurgens11b, Krasnov11a, Katterwe11, Koyama11,  Lin11, Lin11b, Slipchenko11}.

For these mesas there are two regimes where emission occurs \cite{Wang09a, Wang10a}. At moderate input power (``low-bias regime'') there is only little heating ($\lesssim 10$\,mW), and the temperature distribution in the mesa is roughly homogeneous and close to the bath temperature $T_b$. The THz emission observed in this regime presumably can be described by more or less standard Josephson physics. At high input power (``high-bias regime'') a hot spot forms inside the mesa \cite{Wang09a,Wang10a,Guenon10}. The hot spot effectively separates the mesa into a ``cold'' part, which is superconducting, and a hot part, which is in the normal state. The ``cold'' part of the mesa is responsible for THz generation by the Josephson effect. The hot spot also seems to play a role for synchronization \cite{Wang10a,Guenon10,Li12}. It has been found that the size and position of the hot spot, and in consequence also the THz emission, can be manipulated by applying proper bias currents across the mesa \cite{Guenon10}. Thus, in order to understand the mechanism of THz radiation in IJJ mesas, it seems crucially important to develop a detailed understanding of the hot spot formation. The present paper is devoted to this subject.

\begin{figure}
\includegraphics[width=1\columnwidth]{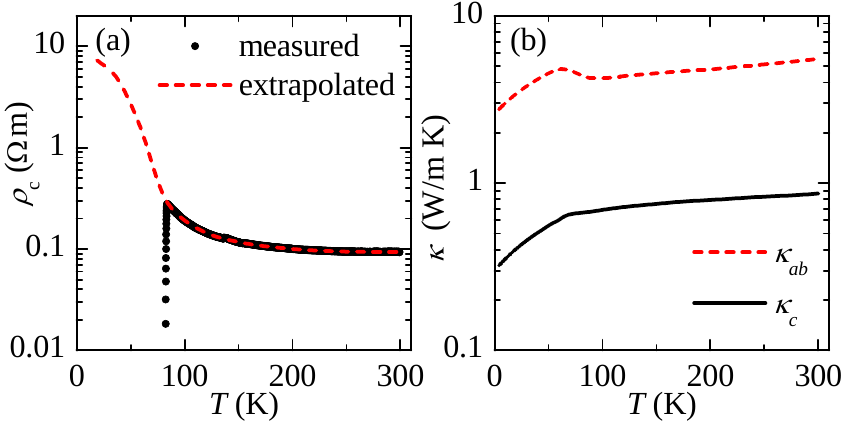}
\caption{(Color online) (a) Temperature dependence of the $c$-axis resistivity $\rho_c$, as measured for a 330 $\times$ 50\,$\mu$m$^2$ wide and 0.7\,$\mu$m thick sample for $T > T_c =$ 83\,K (black circles). For lower $T$, $\rho_c$ has been extrapolated by fitting the IVC, measured at $T_b=20$\,K, using the full 3D heat diffusion equation, cf. Sec. \ref{sec:3D_model}. (b) Temperature dependence of the BSCCO in-plane ($\kappa_{ab}$) and out-of-plane ($\kappa_c$) thermal conductivity \cite{Crommie91}.}
\label{fig:intuitive_model1}
\end{figure}

In a standard superconducting structure (e.g. a thin film) under a strong enough transport current somewhere in the sample the resistance rises from zero to a finite value, leading to local heating and the formation of a hot spot. To obtain THz emission, IJJ stacks are typically biased in a state where all junctions are in their resistive state. Here, the out-of-plane resistance $R_c$ decreases continuously when heating the sample through $T_c$ \cite{Yurgens97, Latyshev99}, \cf Fig. \ref{fig:intuitive_model1} (a). In-plane currents still flow with zero resistance below $T_c$ and with finite resistance above $T_c$.  However, even in the normal state these layers add only a minor contribution to the total voltage across the IJJ stack and thus to the overall power dissipation due to the huge ratio $\rho_c/\rho_{ab} > 10^5$ of the out-of-plane to the in-plane resistivity. It is unlikely that this contribution gives rise to hot spot formation. Also the BSCCO thermal conductance varies relatively weakly with temperature \cite{Crommie91}, cf. Fig. \ref{fig:intuitive_model1} (b). Thus, the above mechanism of hot spot generation does not work and $T_c$ is no longer a peculiar temperature for the thermal balance of the sample.  

There are other ways to create hot domains in systems which may or may not be related to superconductivity \cite{Gurevich87, Volkov69}. In particular, current-voltage characteristics (IVCs) and the thermal breakdown were studied in systems having a resistivity decreasing with increasing temperature (negative-temperature-coefficient resistor) \cite{Spenke35, Lueder36, Spenke36b}. The IVCs of these resistors strongly resemble the IVCs measured for IJJ mesas. Especially, the appearance of a hot domain leads to an abrupt change in differential resistance. The quantity in common, a strongly negative $dR/dT$, is the key to understand hot spot formation in BSCCO mesas.

Recently, Yurgens et al. have simulated the thermal heating and the temperature distribution in BSCCO IJJ mesas \cite{Yurgens10a, Yurgens11}, using a 3D finite-element software \cite{comsol}. In this pioneering work the electrical and thermal properties of the various current carrying and insulating layers were taken into account.
The formation of hot spots observed in \cite{Wang09a} was reproduced qualitatively. However, the occurring phenomena need further study. For example, the IVCs in \cite{Yurgens10a, Yurgens11} have been calculated using a self-consistent procedure based on Newton's law of cooling and Ohm's law and do not exhibit the experimentally observed abrupt changes in differential resistance when the hot spot appears. They resemble much less the experimental curves than the ones calculated in \cite{Spenke35, Lueder36, Spenke36b}. 

A complete study of the Josephson effect in BSCCO mesas in the presence of hot spots is a formidable and unsolved issue. In this paper we are treating experimentally and theoretically hot spot formation in BSCCO mesas. In the theoretical part of our study the presence of the Josephson effect, \ie THz radiation, the formation of electromagnetic standing waves, interactions between hot spots and waves etc. is \textit{not} considered. This approach to hot spot formation seems justified, since the emitted radiation power is 3--4 orders of magnitude lower than the dc input power. It may, however, serve as a zero order approximation towards solving the full problem. In the simulations we derive the electrical current density in the mesa under investigation and thus also the potential difference between top and bottom electrodes, directly generating the IVC for a sample, following \cite{Spenke35, Lueder36, Spenke36b} rather than \cite{Yurgens10a, Yurgens11}.

The paper is organized as follows. In Sec. \ref{sec:intuitivemodel} we consider a simple discrete resistor model to get a basic understanding of the heating phenomena involved. In Sec. \ref{sec:1D_model} a 1D model is discussed which is extended to 3D and realistic sample geometries in Section \ref{sec:3D_model}. The discussions in these sections are based on the thermal and electrical parameters of the BSCCO crystals, as used in experiment. In  Sec. \ref{sec:3D_model} we also address experimental observations, as made in \cite{Wang09a,Guenon10,Wang10a, Li12}. Sec. \ref{sec:conclusion} concludes our work.
%
%
\section{Discrete resistors}
\label{sec:intuitivemodel}
The electrothermal behavior of conducting materials can be investigated by considering the heat balance equation between Joule self-heating $Q(T,\lambda)$ and the heat transfer power $W(T)$ to the coolant $Q(T,\lambda)=W(T)$ \cite{Gurevich87}. Here $\lambda$ is some control parameter (in our case the voltage $V$ across the sample).
\begin{figure}%
\includegraphics[width=1.0\linewidth]{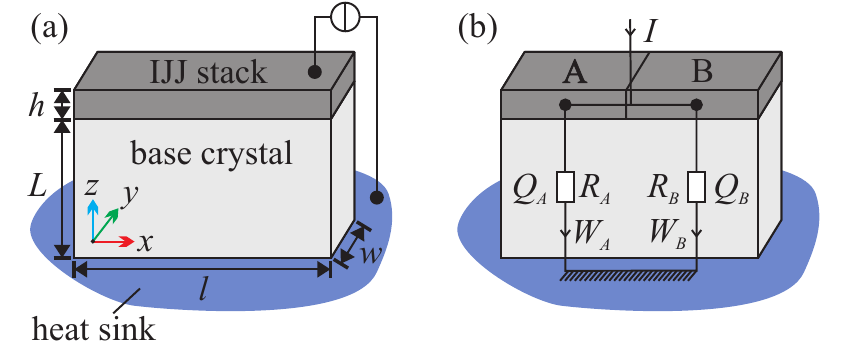}
\caption{(Color online) Discrete approximation for a mesa. (a) Dimensions of mesa and base crystal. (b) The mesa is replaced by two vertically cooled resistors $R_A$ and $R_B$ producing Joule heat $Q_A$ and $Q_B$, which is vertically transported to a thermal bath via heat-transfer powers $W_A$ and $W_B$.}
\label{fig:intuitive_model2}%
\end{figure}
To approach the experimental situation of IJJ mesas we first briefly study the model of two current-biased resistors $R_A(T_A)$ and $R_B(T_A)$ connected in parallel, each representing one half of a mesa of length $l$, width $w$ and height $h$, \cf Fig. \ref{fig:intuitive_model2}. $T_A$ and $T_B$ are the temperatures of these resistors. $R_A$ and $R_B$ shall be equal for $T_A$ = $T_B$. Joule heating is produced via $Q_i=I_iV_i$, where $i = (A,B)$. The total current is $I=I_A+I_B$ and further $V_A$ = $V_B$, \ie we neglect the voltage drop due to in-plane currents. The resistors are thermally connected to a bath (temperature $T_b$), which, at a distance $L$ (the thickness of the base crystal) removes heats $W_A$ and $W_B$ ``vertically'', through the BSCCO out-of-plane thermal conductivity $\kappa_c$. 
%
%
We first assume $T_A=T_B=T$. Then, the IVC of the mesa can be parameterized by $T$, using $Q=W$ \cite{Busch21}:
\begin{equation}
V=\sqrt{R(T)\,W(T-T_b)};\quad
I=\sqrt{\frac{W(T-T_b)}{R(T)}}
\label{eq:parameterization}
\end{equation}
with $ W(T-T_b) = (lw\kappa_c/L) \cdot (T-T_b)$ and $R(T)=(h/l\,w)\cdot\rho_c(T)$. For further calculations we use a constant $\kappa_c = $ 0.6 W/mK. As we want to study the question whether or not the particular $\rho_c(T)$ of our mesas can lead to hot spots we use a temperature dependence which is as close as possible to the experimental situation. Above $T_c$ we obtained $\rho_c(T)$ from the out-of-plane resistance of one of our mesas, cf. solid circles in Fig. \ref{fig:intuitive_model1} (a). Below $T_c$, $\rho_c(T)$ is extrapolated by fitting the measured IVC of the mesa at a bath temperature of 20\,K (see below), using the full 3D heat diffusion equation (dashed line in Fig. \ref{fig:intuitive_model1} (a)). 
 $L$ = 17\,$\mu$m is chosen, which is a typical value for the thickness of the BSCCO base crystal of the samples we want to discuss \cite{Wang09a,Guenon10,Wang10a, Li12}. Length, width and height of the mesa are, respectively, taken to be 330, 50 and 1 $\mu$m, representing sample 1 from \cite{Wang10a}. With these dependencies, the calculated IVC of the mesa is S-shaped and shows a region of negative differential resistance, \cf solid line in Fig. \ref{fig:thermal_branching} (c). In this voltage region thermal bistability can occur, since $W=Q$ holds for more than one value of $T$ \cite{Gurevich87}. In fact, writing $dV/dI = (dV/dT)/(dI/dT) <0$, using Eq. (\ref{eq:parameterization}) and $W \propto (T-T_b)$ one obtains $-(T-T_b)\,(dR/dT)/R  > 1$ as a condition for obtaining negative differential resistance in the IVC and thus the possibility to have  thermal instability \cite{Volkov69}.

We assume for the following, that $I$ and thus $Q$ is increased from zero step-by-step. Fig. \ref{fig:thermal_branching} (a) and (b) show the individual IVCs of resistor A and B respectively, while (c) shows the IVC of the whole mesa. For small $Q$ the temperature is the same in both resistors and they carry the same current. In principle, further increase of $I$ would make the whole mesa pass the point $\delta$ of local maximal voltage $V_{0}$, cf. Fig. \ref{fig:thermal_branching} (c) and enter the unstable \cite{Gurevich87} area of negative differential resistance. This is exemplarily indicated for point $\alpha$ in Fig. \ref{fig:thermal_branching} (c). Here, the two resistors with equal temperature $T_{\alpha}$, would be in states $\alpha_A$ and $\alpha_B$, \cf Fig. \ref{fig:thermal_branching} (a) and (b) respectively. The instability and the constraints of equal voltage and fixed current force the mesa into the state $\beta$, which is composed of state $\beta_A$ with $T_{\beta_A}>T_{\alpha}$ and $\beta_B$ with  $T_{\beta_B}<T_{\alpha}$, \cf Fig. \ref{fig:thermal_branching} (a) and (b) respectively. The combination of $\beta_A$ and $\beta_B$ is the only stable solution. The resulting total IVC of Fig. \ref{fig:thermal_branching} (c) follows the path indicated by the dashed (red) line, differing in voltage from the isothermal case (solid blue line). With increasing $I$, starting from point $\delta$, the points $\beta_A$ and $\beta_B$ ``move'' towards lower voltage. Note that this implies, that the cold resistor becomes colder while the hot resistor keeps increasing its temperature. When $\beta_A$ has reached the minimal voltage, both $\beta_A$ and $\beta_B$ start to move towards larger voltage, \ie also the temperature of the cold part starts to increase. Finally, when $\beta_B$ reaches the voltage $V_0$, $T_A \neq T_B$ becomes impossible and the mesa switches back to the homogeneous solution.
 
\begin{figure}
\includegraphics[width=1\linewidth]{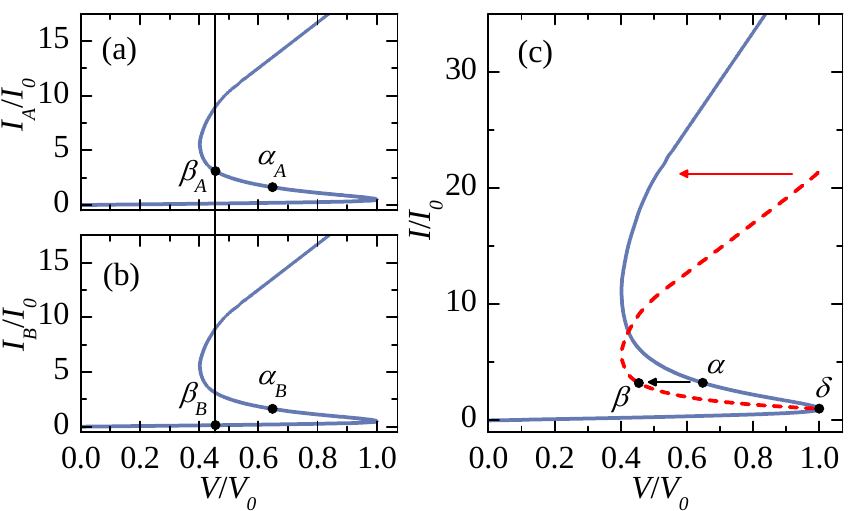}
\caption{(Color online) Hot spot formation in a two-resistor model, cf. Fig. \ref{fig:intuitive_model2}(b). (a) and (b) display the IVCs of the two individual parts A and B, respectively. (c) shows the IVC of the combined system. The axes are normalized to the current (voltage) of the point showing local maximal voltage $V_{0}$. The total current through the mesa at $V_0$ is $I_0$. The bias points indicated by Greek characters are discussed in the text. In (c) for the solid (blue) curve resistors A and B are at the same temperature, while for the dashed (red) curve their temperature differs, corresponding to hot spot formation in the continuous case.}
\label{fig:thermal_branching} 
\end{figure}
The model of two parallel resistors can be extended by several ways: First, an in-plane thermal coupling $W_{AB}$ between resistors A and B may be included. Then, the cold part will cool the hot part and (thermal) differences between A and B will be less severe. This will shift the point, where the homogeneous solution and the solution $T_A \neq T_B$ fork, to higher input power \cite{Spenke35, Spenke36b}. Also, the difference in voltage between the homogeneous and the inhomogeneous solution will be diminished \cite{Spenke35, Spenke36b}. A detailed discussion, however, is out of the scope of this section. In-plane cooling will be taken into account in the subsequent sections. Second, one may allow the two resistors (the area of the ``hot'' and ``cold'' parts) to be unequal and variable in size. Then, one faces a continuous set of solutions. Third, one may consider more than two resistors in parallel. This would be also applicable to the description of arrays of IJJ stacks, which are interesting for obtaining a large THz emission output power. In this scenario, the whole system will tend to a state, where only one of the stacks is hot, while all the others are cold \cite{Spenke36b}.
%
%
\section{1D model}
\label{sec:1D_model}
\begin{figure}
\includegraphics[width=1\linewidth]{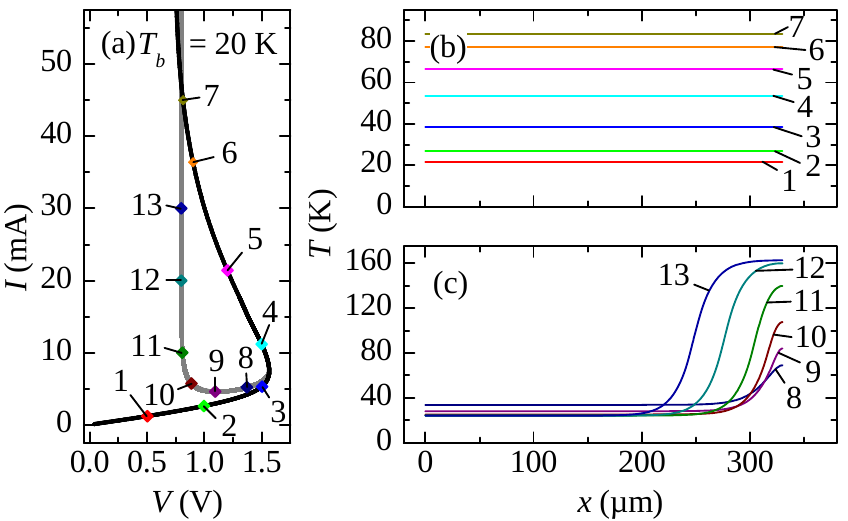}
\includegraphics[width=1\columnwidth]{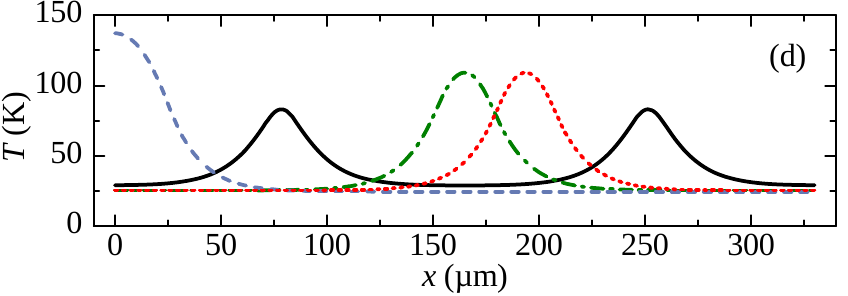}
\caption{(Color online) Simulation results of Eq. (\ref{eq:1D_DGL}) for $T_b=20$\,K and $L=19$\,$\mu$m. Other mesa dimensions are listed in Sec. \ref{sec:intuitivemodel}. Graph (a) shows the IVC for the homogeneous solution (black curve) and a solution showing a hot spot on the right mesa end (gray curve). $T(x)$-profiles are displayed in (b) for the homogeneous case and in (c) for the hot spot case. The numbers indicate the bias points on the IVC. In (d) $T(x)$-profiles, obtained from different $T_i(x)$, are shown for solutions with $I=9.5$ mA, exhibiting various shape and positioning of the hot spot.}
\label{fig:iv001_1D} 
\end{figure}

In this section we consider a 1D continuous model to find the temperature distribution in the mesa for the simplest continuous case, still treating hot spot formation from a generic point of view. That is, we assume a thin (along $z$) and narrow (along $y$) mesa, neglecting $T$-variations along $z$- and $y$-direction in the mesa (see \cite{Spenke36b, Gurevich87} for details). Then, $T=T(x)$ is defined by the heat diffusion equation:
\begin{equation}
\begin{split}
-h \frac{d}{dx}\left[\kappa_{ab}\left(T\right)\frac{d}{dx} T\right]+ \frac{\kappa_{c}\left(T\right)}{L}\, (T-T_b) = \frac{V^2}{\rho_c\left(T\right)\, h}
\end{split}
\label{eq:1D_DGL}
\end{equation}
The first term describes the thermal diffusion in $x$-direction and the second one the cooling due to the base crystal with the coefficient $\kappa_c/L$ regulating its strength. The third term represents Joule heating. The sample dimensions $L$, $h$, $l$ and $w$ are defined in Fig. \ref{fig:intuitive_model2} (a). We use $L$ = 19\,$\mu$m, $h$ = 1\,$\mu$m, $l$ = 330\,$\mu$m, $w$ = 50\,$\mu$m and $\kappa_{ab}$, $\kappa_{c}$ and $\rho_c$ as in Figs. \ref{fig:intuitive_model1} (a), (b). The boundary conditions are chosen to be $dT/dx\left(x=0\right)=dT/dx\left(x=l\right)=0$. These boundary conditions neglect edge cooling. To solve Eq. (\ref{eq:1D_DGL}) for a given current $I$ we use $V = I h/\int\rho_c^{-1}\left(T\right)dxdy$. We numerically solve Eq. (\ref{eq:1D_DGL}) using finite element analysis \cite{comsol}. Note that there is always a homogeneous solution. To find a nontrivial $T\left(x\right)$, a proper initial function $T_i\left(x\right)$ has to be used. A calculated IVC for $T_b=20$\, K is shown in Fig. \ref{fig:iv001_1D} (a). It resembles the shape of the IVC of the two-resistor model, \cf Fig. \ref{fig:thermal_branching}.
Figure \ref{fig:iv001_1D} (b) shows the temperature in the mesa for the homogeneous solution at the bias points indicated in (a). One notes that the mesa temperature is below $T_c$ up to quite high currents $\sim$ 40\,mA.
Figure \ref{fig:iv001_1D} (c) shows solutions for the bias points indicated in the IVC, when a hot spot has formed. Here, the temperature in the hot part rises rapidly to temperatures well above $T_c$, while the temperature of the cold part is near $T_b$ = 20\,K. Also, one observes, that the hot part grows in size when $I$ is increased. Further note, that in the presence of a hot spot the temperature $T_{cold}$ of the cold part is below the temperature of the homogeneous solution for the same value of $Q$;
$T_{cold}$ decreases with increasing $Q$, and finally converges against a limiting value. The strength of the deviation of the temperature profile from the homogeneous solution directly correlates with the strength of branching in the IVC. In the depicted case the branching is very strong, which is due to a small ratio of the in-plane to the out-of-plane thermal coupling, \cf first and second term in Eq.  (\ref{eq:1D_DGL}).

For a given $I$ the hot spots presented in Fig. \ref{fig:iv001_1D} (c) are not the only possible solutions to Eq. (\ref{eq:1D_DGL}) \cite{solutionprocess}. For symmetry reasons also the mirrored solution exists, as well as solutions with the hot spot near and in the center of the mesa, \cf Fig. \ref{fig:iv001_1D} (d).
In the IVC the different solutions slightly differ in $V$ and can be traced over some range in $I$. Thus, the IVC consists of several branches distinguishing specific kinds of hot domains. Experimentally, in some cases, hot spot formation in different places of the mesa has been detected by low-temperature scanning-laser microscopy (LTSLM). However, usually a specific configuration is much more stable than the others, presumably due to inhomogeneities like attached wires. In the calculations, also solutions with more than one hot domain \cite{severalhotdomains} can be found, \cf Fig. \ref{fig:iv001_1D} (d).
However, this has not been observed in any of our LTSLM measurements. It is argued in \cite{Spenke36b}, that such a state is very unstable and will not occur, since the sample can be seen as a parallel circuit of several discrete parts with small thermal coupling between them, \cf section \ref{sec:intuitivemodel}.
%
%
\section{3D model}
\label{sec:3D_model}
\begin{figure}
\includegraphics[width=1\linewidth]{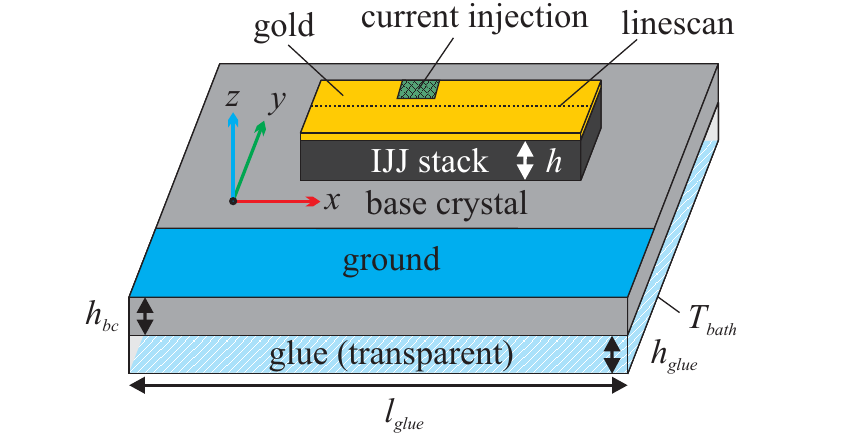}
\caption{(Color online) Model geometry of the mesa.}
\label{fig:3D_model_geometry} 
\end{figure}
In this section we address hot spot formation in 3D. The goal is to quantitatively compare our experimental observations with the numerical simulations, using the same code \cite{comsol} as in \cite{Yurgens10a, Yurgens11}. Similarly, we also include various electrically and thermally conducting and insulating layers that are in contact with our BSCCO mesas. The electrical, thermal and geometrical parameters used for the calculations are as close as possible to the experimental situation. The geometry used is still somewhat simplified compared to the real samples but should allow to capture the relevant physics. Figure \ref{fig:3D_model_geometry} depicts the model. The substrate is omitted and a boundary condition $T_b$ = const. is applied to the bottom surface of the glue layer, representing the thermal bath. This simplification can be done with very little impact on the results for the mesa, since the thermal conductivity of the substrate (e.g sapphire) is by far better than that of the glue layer. The geometric dimensions of the mesa, the thicknesses of the glue layers (10 -- 30 $\mu$m) and of the gold coatings ($h_{Au} \approx$\, 30 nm) were roughly chosen as in the real samples. The base crystal's lateral size is typically of the order of 1 mm, while its thickness $h_{bc}$ may vary from about ten to several hundred $\mu$m, strongly depending on the fabrication process. The current leads are simply represented by boundary conditions on the surfaces of either the gold layer on the mesa or on the base crystal. The current $I$ is injected through a 20 $\times$ 10\,$\mu$m$^2$ rectangle and the current sink is defined as a ground of large area (roughly 0.3 mm$^2$), \cf Fig. \ref{fig:3D_model_geometry}. The voltage across the mesa is obtained as the potential difference between the two electrodes. 

The equation to be solved is \cite{Yurgens11}:
\begin{equation}
-\nabla\left[\kappa\left(T\left(\bf{r}\right)\right)\:\nabla T\left(\bf{r}\right)\right]=\rho\left(T\left(\bf{r}\right)\right)\:\bf{j}^2\left(\bf{r}\right),
\label{eq:3D_DGL}
\end{equation}
where $\rho$ and $\kappa$ are the resistivity and thermal conductivity tensor, respectively, and \textbf{r} is the spatial coordinate. Unlike the mesa, the base crystal is not always in the resistive state. We model its resistance  by using the $\rho_{c}$ vs. $T$ data indicated by solid circles in Fig. \ref{fig:intuitive_model1} (b). The in-plane resistivity $\rho_{ab}$ is the same for both mesa and base crystal; we use the same $T$ dependence as in \cite{Yurgens11}. The thermal conductivity for BSCCO is used from \cite{Crommie91}, \cf Fig. \ref{fig:intuitive_model1} (a). Thermal and electrical conductivity for a 30 nm thick Au film are adopted from \cite{Chopra63}. For the thermal conductivity $\kappa_{\rm{glue}}$ of the glue between the BSCCO base crystal and the substrate to first order we use the polyimide data of \cite{NIST}. Since our glue might have slightly different properties we in addition multiply $\kappa_{\rm{glue}}$ with a factor $n_{\rm{glue}}$, which we fit by adjusting the calculated IVC to the measured one. 
%

The base crystal introduces an effective side-cooling of the mesa, which in general makes a solution showing variation in $x$ and $y$ direction (with or without hot spots) favorable. Indeed, in contrast to the one-dimensional calculations, hot spot solutions appeared basically by themselves, \ie it was not necessary to find them by choosing a proper initial condition. The side-cooling leads to an elliptic shape of the hot spot (for rectangular shaped mesas). Also, the hot spot is not limited to the mesa itself anymore, but may extend significantly in lateral direction into the base crystal (see below). This is exactly what has been found experimentally \cite{Guenon10}. The same occurs in $z$-direction, as has been discussed in \cite{Yurgens11}.

%
\begin{figure}
\includegraphics[width=1\linewidth]{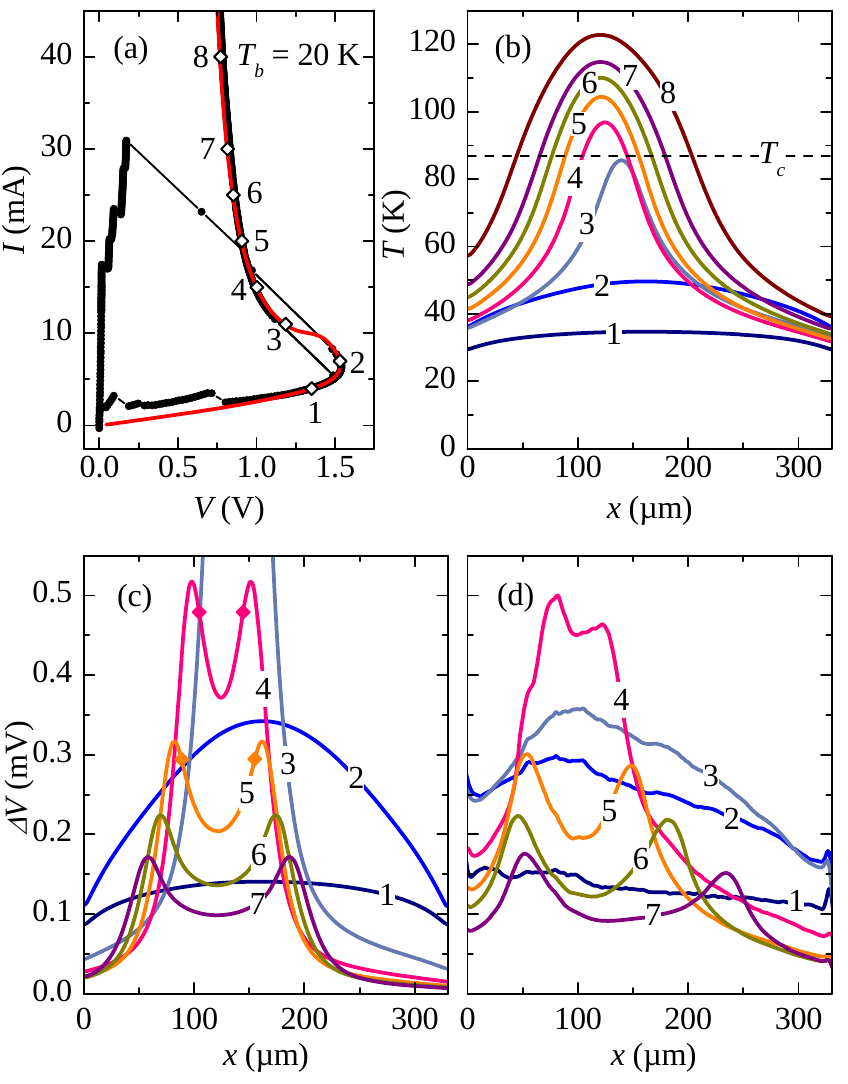}
\caption{(Color online) Comparison of 3D simulation and experimental data for sample 1 from \cite{Wang10a} at $T_b$ =\,20 K. (a) shows the measured (black, solid circles) and simulated (red solid line) IVCs. In (b) simulated $T(x)$-profiles along the dashed line indicated in Fig. \ref{fig:3D_model_geometry} at $z=0.5\, h$ are shown. The diamonds in (a) indicate the corresponding bias points. The calculated and measured $\Delta V(x)$ are shown in (c) and (d), respectively.  Diamonds in (c) indicate the $x$-position where  $T=T_c$.}
\label{fig:No1-20K-compare-DeltaV} 
\end{figure}
We investigate sample 1 from \cite{Wang10a}. The electrical and thermal parameters of this sample have already been used in the previous sections. We have further used the parameters: $h_{bc}=$\,40 $\mu$m, $h_{glue}=$\,25 $\mu$m, $l_{glue}=$\,1 mm and $n_{glue}=1.95$. Figure \ref{fig:No1-20K-compare-DeltaV} (a) compares the measured IVC with the calculated one for $T_b=20$\, K. The good agreement stems from the fact that we have adjusted $\sigma_c\left(T\right)$ below $T_c$ and $n_{glue}$ to match this curve. The simulated $T(x)$-profiles, calculated along the dashed line in Fig. \ref{fig:3D_model_geometry} at $z=0.5\,h$, are shown in Fig. \ref{fig:No1-20K-compare-DeltaV} (b) for the bias points indicated in Fig. \ref{fig:No1-20K-compare-DeltaV} (a). They show an almost constant temperature in the low bias regime, whereas for increasing current the hot spot forms by the growth of a buckling in the $T(x)$-profile (compare curves 2 and 3 in Fig. \ref{fig:No1-20K-compare-DeltaV} (b)). Further increasing $I$ and thus $Q$ leads to a growth in diameter and maximal temperature of the hot spot (curves 4 to 8). Note that $T_c$, indicated by the horizontal dashed line, can be significantly exceeded in the center of the hot spots, confirming the results in \cite{Yurgens11}.

%
We next want to provide a quantitative comparison between the hot spot signals observed in LTSLM \cite{Wang09a,Guenon10,Wang10a} and the calculated temperature distributions for this sample. In LTSLM the laser spot at position ($x_0$,$y_0$) causes a maximum temperature rise $\Delta T \sim$ 1-3 K, depending on the laser power. In turn there is a change $\Delta V (x_0, y_0)$ in the voltage $V$ across a sample. One often has a response which partially arises from the reduction of the Josephson critical current density and partially from the change in resistance, see e.g. \cite{Guerlich10}. However, if $dR_c(T)/dT$ dominates the thermal physics, $\Delta V (x_0, y_0)$ can be treated as in \cite{Werner11} yielding
\begin{equation}
\Delta V\left(x_0,y_0\right)\approx\frac{-IR^{2}_{eff}\Delta TA_{L}}{h}\frac{d\sigma_c}{dT}\left(T\left(x_0,y_0\right)\right).
\label{eq:deltaV}
\end{equation}
$R_{eff} = V/I$ is the (ohmic) sample resistance at a given $I$ and $A_L$ is the effective area warmed up by the laser (some $\mu$m$^2$). $d\sigma_c/dT\left(T\left(x_0,y_0\right)\right)$ denotes the temperature derivative of the c-axis electrical conductivity. The calculated and measured $\Delta V$, taken at various bias points indicated in Fig. \ref{fig:No1-20K-compare-DeltaV} (a), are shown in Fig. \ref{fig:No1-20K-compare-DeltaV} (c) and (d), respectively. For the simulations we have used $\Delta T\cdot A_L$ = 56\,K$\mu$m$^2$. The value makes sense, since we expect a temperature rise $\Delta T \sim$ 2\,K and $A_L\sim 25$\,$\mu$m$^2$ for the samples we discuss here. The calculated curves agree reasonably well with the measurements, although differences occur at low bias and near the hot spot nucleation point. Particularly, for the bias points 1 and 2, the simulation yields a parabolic shape of $\Delta V$, while the experimental data are shaped less regular. Note, however, that in these regions the Josephson currents, which are neglected in our analysis, may play a major role. For curve 3 in the simulation hot spot formation has already occurred, while in experiment the mesa is close to the nucleation point but still undercritical. For a bias well above the hot spot nucleation point theoretical curves and experimental data agree well. Specifically the double hump feature in $\Delta V\left(x\right)$ is reproduced correctly in the simulations. The local temperature at the maxima in $\Delta V$ corresponds to the temperature $T^{*}\approx 80$\,K, for which $d\sigma_c/dT$ is maximum, \cf Eq. (\ref{eq:deltaV}). Between the two $\Delta V$ maxima, $T>T^{*}$. By coincidence, $T^{*}\approx T_c$; the diamonds in Fig. \ref{fig:No1-20K-compare-DeltaV} (c) indicate the locations for which $T=T_c$. Thus, the border between superconducting and non-superconducting parts, which is important for THz emission, can be approximately identified by the position of the humps.

\begin{figure}
\includegraphics[width=1\linewidth]{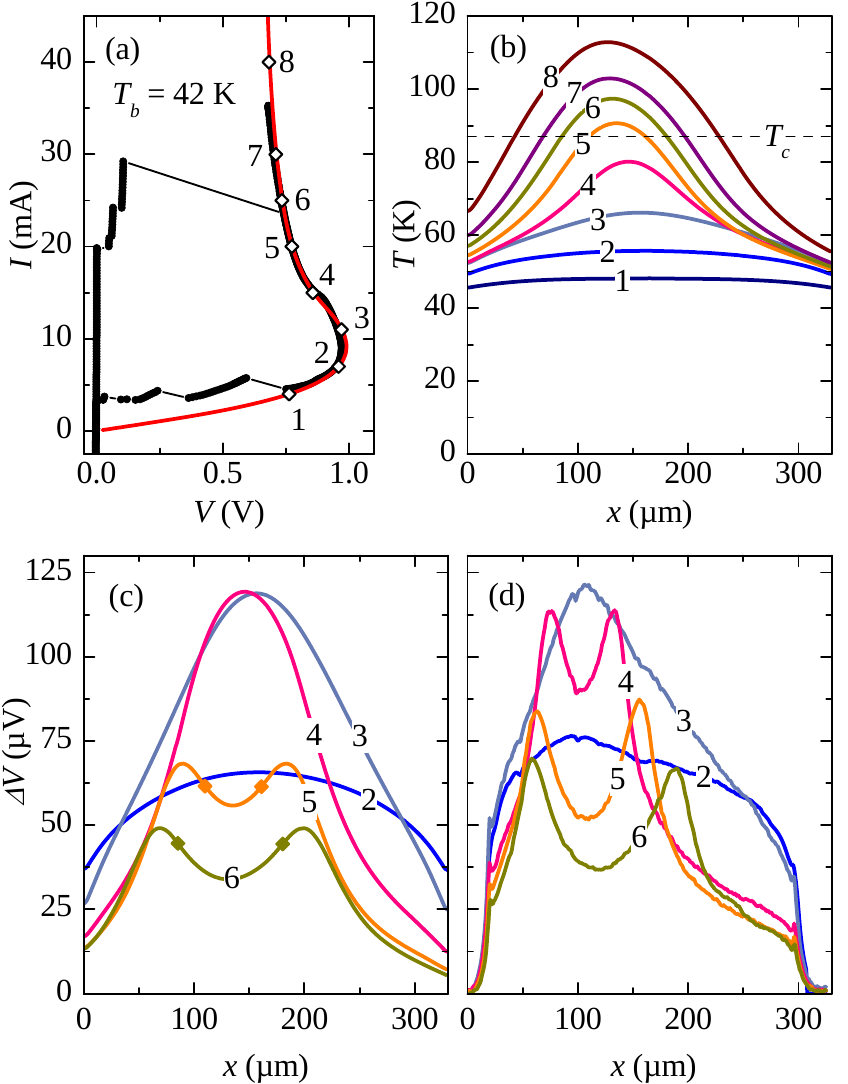}
\caption{(Color online) Comparison of 3D simulation and experimental data for sample 1 from \cite{Wang10a} at $T_b$ = 42\,K. (a) shows the measured (black, solid circles) and simulated (red solid curve) IVC. In (b) simulated $T(x)$-profiles along the dashed line indicated in Fig. \ref{fig:3D_model_geometry} at $z=0.5\, h$ are shown. The diamonds in (a) indicate the corresponding bias points. The calculated and measured $\Delta V(x)$ are shown in (c) and (d), respectively.  Diamonds in (c) indicate the $x$-position where  $T=T_c$.}
\label{fig:No1-42K-compare-DeltaV} 
\end{figure}
We next investigate the dependence of hot spot formation on $T_b$. Figure \ref{fig:No1-42K-compare-DeltaV} shows a similar set of data as Fig. \ref{fig:No1-20K-compare-DeltaV}, but for $T_b=$ 42 K. Here, $n_{glue}=3.5$ has been chosen. The transition region between the hot and cold domain is less steep than for $T_b$ = 20\,K. Also, the nucleation point of the hot spot has moved to higher currents (10 mA for $T_b$ = 20\,K and 14 mA for $T_b$ = 42\,K) and the back-bending of the IVC has decreased. These effects arise from the fact that the $xy$-plane thermal coupling has increased relative to the out-of-plane thermal coupling \cite{Spenke36b}, \cf Sec. \ref{sec:1D_model}. Note, that this also means for equal input power, that the hot domain reaches higher and the cold domain reaches lower temperatures for $T_b$ = 20\,K as for $T_b$ = 42\,K. Figures \ref{fig:No1-42K-compare-DeltaV} (c) and (d) respectively show the calculated and measured LTSLM profiles. As for \ref{fig:No1-20K-compare-DeltaV} (c) and (d) the agreement between experimental data and simulations is reasonable, except for the bias point where hot spot formation sets in (curve 4). For the calculations we have used  $\Delta T \cdot A_L$ = 16\,K$\mu$m$^2$, which is by a factor of 3 lower than for the case of $T_b$ = 20\,K. This is attributed to a reduced incident laser power, which had been readjusted for every measurement.

\begin{figure}
\includegraphics[width=1\linewidth]{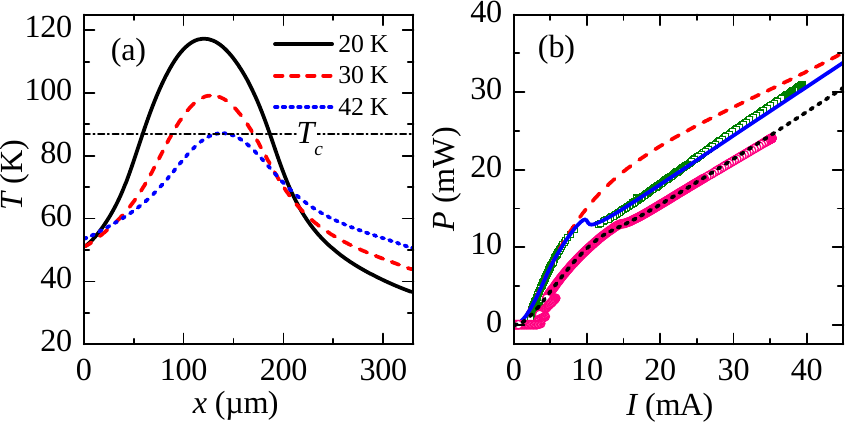}
\caption{(Color online) 
(a) Simulated $T(x)$-profiles along the dashed line indicated in Fig. \ref{fig:3D_model_geometry} at $z=0.5\, h$ for three different values of $T_b$ at constant $V=0.8$ V. The dc power $P = IV$ is 26.4, 18.2 and 14.4 mW for $T_b=$ 20, 30 and 42 K, respectively. (b) $P$ vs. $I$ at 20 K for measurement (green squares), simulation with homogeneous $T$ (red dashed line) and hot spot formation (blue, solid line). The pink circles depict experimental data and the black, short-dashed line simulated values with hot spot formation at 42 K.}
\label{fig:PvsI} 
\end{figure}
In Fig. \ref{fig:PvsI} (a), we show the $T\left(x\right)$-profile, calculated along the dashed line in Fig. \ref{fig:3D_model_geometry} at $z = 0.5\,h$, for 3 values of $T_b$. For all curves, $V$ = 0.8\,V. This condition has been motivated by measurements of the linewidth $\Delta f$ of THz radiation \cite{Li12}. Here, for $\Delta f$ vs. $T_b$, taken at a fixed emission frequency (corresponding to $V$ = const. for a fixed number of oscillating IJJs) the dependence $\Delta f \propto T_b^{-4}$ has been found unexpectedly. We are interested in the question whether or not corresponding changes with $T_b$ can be seen in the $T$ distribution in the mesa. Figure \ref{fig:PvsI} (a) shows, that the peak temperature in the mesa is higher at low $T_b$ than at high $T_b$, while the coldest temperatures, reached at the right edge of the mesa, behave oppositely. Thus, thermal gradients at low $T_b$ are stronger than at high $T_b$. However, this effect roughly changes linearly with $T_b$ and presumably cannot explain the $\Delta f \propto T_b^{-4}$ dependence.

Figure \ref{fig:PvsI} (b) compares for two values of $T_b$ the measured and simulated DC power $P = IV$ as a function of $I$. One observes two regimes, each with a roughly constant slope. The first -- low-bias -- regime has no hot spot and, for $T_b =20$\,K, spans from 0 to 10 mA (14 mA for 42 K), whereas the second -- high-bias -- regime has a hot spot and begins at 10 mA (14 mA for 42 K). Interestingly, at the intersection of these two regimes the maximum temperature in the mesa has reached the temperature fulfilling $d\sigma_c/dT=0$. This point also corresponds to the kink in the IVC, observed for several mesas. Note that the calculation for homogeneous T (red dashed curve) shows no such kink. A plot like this may thus be helpful to distinguish in an experiment, whether or not one has reached the regime with hot spots.

\begin{figure}%
\includegraphics[width=0.491\linewidth]{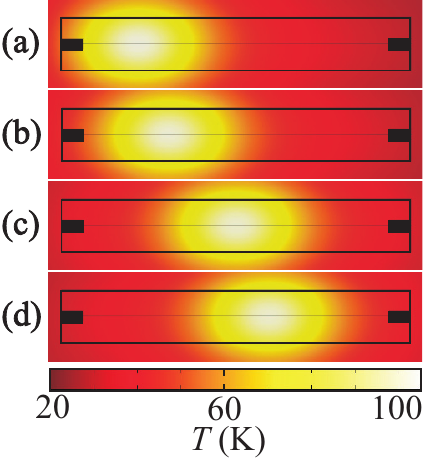}
\includegraphics[width=0.491\linewidth]{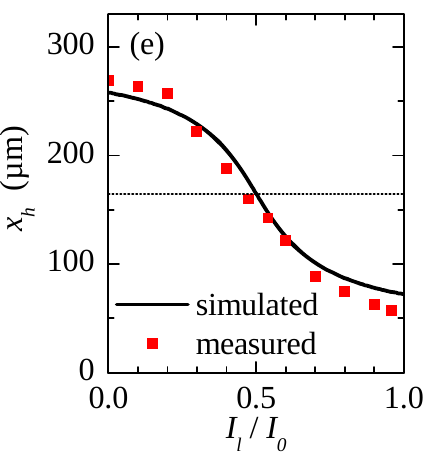}
\caption{(Color online) (a)--(d) Surface plot of hot spot solutions, obtained by Eq. (\ref{eq:3D_DGL}) for a mesa with two current injection points, indicated by black rectangles. The sum $I_0$ of the currents through the left ($I_l$) and right ($I_r$) injection points has been kept constant, and for the ratio $I_l/I_0$ values of (a) 1, (b) 0.7, (c) 0.5 and (d) 0.425 have been used.  Graph (e) shows the center position $x_h$ of the hot spot vs. current injection ratio for simulated and measured data.}
\label{fig:movehotdomain}%
\end{figure}

The last issue we want to address is the correlation between the point of current injection and the location, where the hot spot is established. Typically, in experiment the appearance of the hot spot was close to, but not exactly at the bond wire to the mesa surface \cite{Guenon10}. We see the same effect in our simulations, Fig. \ref{fig:movehotdomain} (a) illustrates this for a situation, where the current is injected from the left. Here, the side-cooling prevents the hot spot from nucleating at the very left end of the mesa, resulting in a positioning of the hot spot at several \micron right of the current injection point.
Further, it has been shown that by using two injectors located on opposite sides of the mesas the hot spot can be moved by changing the ratio of currents through these injectors \cite{Guenon10}. Figure \ref{fig:movehotdomain} (a)--(d) show a sequence of calculations where the ratio between injection currents through the left (current $I_l$) and right (current $I_r$) was varied, keeping the sum of the currents $I_0$ constant. We used ratios $I_L/I_0$ of, respectively, 1, 0.7, 0.5 and 0.425. As one can see, the hot spot indeed can be moved continuously, as in experiment. In Fig. \ref{fig:movehotdomain} (e) we have plotted the center position of the hot spot as a function of $I_l/I_0$. Experimental data are shown by (red) squares and theoretical data by the (black) solid line. The agreement is reasonable, showing that this effect can be essentially understood from the thermal calculations presented in this paper.

Finally, we briefly mention that also two other geometries discussed in \cite{Guenon10} can be reproduced very well in the 3D simulation -- a disk shaped mesa and a mesa of Y shape, where the hot spot forms at the intersection of the three lines, although the bias current injection point was at the foot of the Y. 
%
%
\section{conclusion}
\label{sec:conclusion}
In conclusion, we have investigated experimentally and numerically the temperature profiles and hot spot formation in IJJ mesas. We have shown, that the hot spots dominantly arise because of the strongly negative temperature coefficient of the out-of-plane resistance of the mesas. This mechanism is different from the more conventional hot spot formation in superconductors and, in particular, allows for hot spots with a maximum temperature below as well as above the transition temperature $T_c$. 
We have given -- in the frame of what available data allow -- a quantitative comparison between simulation and experiment, showing reasonable agreement. Numerous effects observed in previous papers on hot spot formation in intrinsic Josephson junction stacks \cite{Wang10a, Wang09a, Guenon10} are reproduced by the simulations, making us confident that the description given in this paper captures the essential physics, except for the interplay of hot spots and THz waves. Resolving this issue is a task for the future.

%
%
\acknowledgments
We gratefully acknowledge financial support by the JST/DFG strategic Japanese-German International Cooperative Program and the German Israeli Foundation (Grant No. G--967--126.14/2007). S.G. acknowledges support by AFOSR.
%
%
\bibliography{etd_refs}

\begin{thebibliography}{72}
\bibitem[{sol()}]{solutionprocess}
\bibinfo{note}{Generally, to find a given solution and to trace it out over
  some current range, the current was first ramped from zero to high currents
  with a homogeneous temperature distribution. In order to get solutions with
  domain formation, a proper initial condition for the temperature profile of
  the ETD solution was made and, if successful, the current was increased or
  decreased in small steps, resulting in a branch with domain formation.}
\bibitem[{sev()}]{severalhotdomains}
\bibinfo{note}{In other systems, stable configurations with few or even many
  hot spots in a sequence were considered for the case, when the out-of-plane
  resistivity is much higher than the in-plane resistivity \cite{Gurevich87,
  Akhmetov83, Akhmetov85}. In this case the high out-of-plane resistivity
  results in an effective self-consistent shunt that is stabilizing the hot
  spot structure.}
\end{thebibliography}
\end{document}